\documentclass[aps,prl,twocolumn, showpacs]{revtex4}

\def\beq{\begin{equation}}
\def\eeq{\end{equation}}
\def\bea{\begin{eqnarray}}
\def\eea{\end{eqnarray}}
\def\nn{\nonumber}
\def\sss{\scriptscriptstyle}
\makeatletter
\newcommand{\contraction}[5][1ex]{%
  \mathchoice
    {\contraction@\displaystyle{#2}{#3}{#4}{#5}{#1}}%
    {\contraction@\textstyle{#2}{#3}{#4}{#5}{#1}}%
    {\contraction@\scriptstyle{#2}{#3}{#4}{#5}{#1}}%
    {\contraction@\scriptscriptstyle{#2}{#3}{#4}{#5}{#1}}}%
\newcommand{\contraction@}[6]{%
  \setbox0=\hbox{$#1#2$}%
  \setbox2=\hbox{$#1#3$}%
  \setbox4=\hbox{$#1#4$}%
  \setbox6=\hbox{$#1#5$}%
  \dimen0=\wd2%
  \advance\dimen0 by \wd6%
  \divide\dimen0 by 2%
  \advance\dimen0 by \wd4%
  \vbox{%
    \hbox to 0pt{%
      \kern \wd0%
      \kern 0.4\wd2%pour deplacer horizontalement!!!!!
      \contraction@@{\dimen0}{#6}%
      \hss}%
    \vskip 0.8ex%pour deplacer verticalement!!!!!
    \vskip\ht2}}

\newcommand{\contraction@@}[3][0.06em]{%
  \hbox{%
    \vrule width #1 height 0pt depth #3%
    \vrule width #2 height 0pt depth #1%
    \vrule width #1 height 0pt depth #3%
    \relax}}
\makeatother
\def\npb#1#2#3{{ Nucl.\ Phys. B} {\bf #1}, #3 (#2)}
\def\plb#1#2#3{{ Phys.\ Lett. B} {\bf #1}, #3 (#2)}
\def\prd#1#2#3{{ Phys.\ Rev. D} {\bf #1}, #3 (#2)}
\def\newprd#1#2#3{{ Phys.\ Rev. D} {\bf #1}, #3 (#2)}
\def\prl#1#2#3{{ Phys.\ Rev.\ Lett.} {\bf #1}, #3 (#2)}

\begin{document}

\preprint{UdeM-GPP-TH-05-133} 
\title{Relative Sizes of Diagrams in $B \to \pi \pi, \pi K$ Decays}
\author{Maxime Imbeault}
\email{maxime.imbeault@umontreal.ca} 
\affiliation{Laboratoire Ren\'e J.-A. L\'evesque, Universit\'e de
Montr\'eal, C.P. 6128, succ.~centre-ville, Montr\'eal, Qu\'ebec,
Canada H3C 3J7}
\date{\today}
\begin{abstract}
We show that the neglect of the $(V-A)\times (V+A)$ pieces of the
electroweak penguin (EWP) amplitudes in the effective hamiltonian (the
Wilson coefficients are very small) allows one to calculate the
relative size of some tree and EWP diagrams in $B \to \pi \pi$ and
$B\to \pi K$ decays. For both decay classes, tree and EWP amplitudes
are related using only isospin. In $B\to \pi \pi$, the ratio $C/T$
is calculated using isospin alone; in $B\to \pi K$ it is found using
flavor SU(3) symmetry. These results are obtained by computing
explicitly all Wick contractions of all effective operators.
Relations among these contractions are found using Fierz identities
and final-state symmetry arguments.
\end{abstract}
\maketitle 

A very useful way to parametrize $B$ decays is through the use of
diagrams \cite{GHLR}. The size of these diagrams is a priori unknown
and can only be estimated using theoretical input. In this letter, we
show that, in fact, using simple field-theoretical tools, it is
possible to compute the relative size of certain diagrams. (Note that
the present paper focuses principally on explanations and results.
For the full calculation see Ref.~\cite{avenir}.)

The starting point is the effective hamiltonian. The Wilson
coefficients of the $(V-A)\times (V+A)$ pieces of the electroweak
penguin (EWP) operators are tiny, and can be neglected \cite{NR}.
Thus, the operator form of tree and EWP amplitudes is identical: both
are $(V-A)\times (V-A)$. It is already known that this fact leads to
some relations among certain diagrams \cite{NR, GPY}, but flavor SU(3)
symmetry is always required, especially for $B\to \pi K$ decays. Our
approach is different, and we (surprisingly) prove that some relations
between tree and EWP diagrams can be obtained with isospin alone, even
for $B\to \pi K$ decays. We simply ``sandwich'' all effective
operators of the effective hamiltonian between initial and final
states, and apply the basic rules of quantum field theory by summing
over all possible Wick contractions. In so doing it is possible to
write tree and EWP diagrams in term of these Wick contractions. Then,
using Fierz identities and final-state symmetry arguments (isospin is
assumed), many Wick contractions, or diagrams, can be related. This
allows us to calculate the ratio of tree and EWP diagrams, including
their color suppression. Further relations are obtained by adding
flavor SU(3) symmetry. We find many new results and reproduce some
others, in agreement with those found in published papers \cite{NR,
GPY}. We explain why the new relations, using only isospin, could not
be obtained by previous SU(3) analysis.

For $B\to \pi \pi$ decays the effective hamiltonian is \cite{BBNS}.
\beq 
H_{\sss eff} = {G_F \over \protect \sqrt{2}}
\left(\sum_{p=u,c}\lambda_p^{(d)} (c_1 O^p_1 + c_2 O^p_2) -
\lambda_t^{(d)} \sum_{i=3}^{10} c_i O_i \right)~,
\label{Heff}
\eeq
where $\lambda_i^{(d)}=V_{ib}V^*_{id}$, and the tree and EWP operators
are respectively (we neglect $Q_7$ and $Q_8$, the $(V-A)\times (V+A)$
EWP's)
\bea
Q_1^p &=& (\bar p b)_{\sss V-A} (\bar d p)_{\sss V-A}~,\nn\\ Q_2^p &=&
(\bar p_i b_j)_{\sss V-A} (\bar d_j p_i)_{\sss V-A}~,\nn\\ Q_9 &=&
\frac{3}{2} (\bar d b)_{\sss V-A} \sum_{q} e_{q} (\bar q q)_{\sss V-
A}~,\nn\\
Q_{10} &=& \frac{3}{2} (\bar d_i b_j)_{\sss V-A} \sum_{q} e_{q} (\bar
q_j q_i)_{\sss V- A}~,
\eea  
with $q=u,d,s,c,b$. For $B\to \pi K$, the non-summed $d$ quarks are
replaced by an $s$ quark. Factors of $G_F/\sqrt{2}$ are omitted for
the remainder of this paper.

When sandwiching these operators between initial and final states, all
terms have the form
\beq
\langle \bar q_1 q_2 \bar q_3 q_4 | \bar q_5 b \bar q_6 q_7 | \bar q_8
b \rangle~,	
\eeq
where the $q_i$'s are $u$, $d$ or $s$ quarks, $\bar q_1 q_2$ and $\bar
q_3 q_4$ are $\pi$ or $K$ mesons and $\bar q_8 b$ is a $B$ meson
(Dirac and color structures are omitted for notational convenience).
Applying the basic rules of quantum field theory, we must sum over all
possible Wick contractions of all operators. There are 24 possible
contractions and we assign them labels from $A$ to $X$ (see
Table~\ref{contractions}).

\begin{table}
\caption{The various Wick contractions for the decay $\bar q_8 b \to
\bar q_1 q_2 \bar q_3 q_4$. Any contraction not listed is symmetric to
one listed by the exchange $\bar q_1 q_2 \leftrightarrow \bar q_3
q_4$.}
\begin{center}
\begin{tabular}{cc}
$A= { \contraction[1ex]{\langle \bar q_1 }{q_2 }{}{\bar q_3 }
\contraction[1ex]{\langle \bar q_1 q_2 \bar q_3 }{q_4 }{| \bar q_5 b
}{\bar q_6 }
\contraction[2ex]{\langle \bar q_1 q_2 \bar q_3 q_4 | }{\bar q_5 }{b
\bar q_6 }{q_7 }
\contraction[3ex]{\langle }{\bar q_1 }{q_2 \bar q_3 q_4 | \bar q_5 b
\bar q_6 q_7 | }{\bar q_8}
\langle \bar q_1 q_2 \bar q_3 q_4 | \bar q_5 b \bar q_6 q_7 | \bar q_8
b \rangle }$ &
$B= { \contraction[1ex]{\langle \bar q_1 }{q_2 }{}{\bar q_3 }
\contraction[1ex]{\langle \bar q_1 q_2 \bar q_3 }{q_4 }{| \bar q_5 b
}{\bar q_6 }
\contraction[2ex]{\langle }{\bar q_1 }{ q_2 \bar q_3 q_4 | \bar q_5 b
\bar q_6 }{q_7 }
\contraction[3ex]{\langle \bar q_1 q_2 \bar q_3 q_4 | }{\bar q_5 }{b
\bar q_6 q_7 | }{\bar q_8 }
\langle \bar q_1 q_2 \bar q_3 q_4 | \bar q_5 b \bar q_6 q_7 | \bar q_8
b \rangle }$ \\
$C= { \contraction[1ex]{\langle }{\bar q_1 }{}{q_2 }
\contraction[1ex]{\langle \bar q_1 q_2 \bar q_3 q_4 | }{\bar q_5 }{b
\bar q_6 }{q_7 }
\contraction[2ex]{\langle \bar q_1 q_2 \bar q_3 }{q_4 }{| \bar q_5 b
}{\bar q_6 }
\contraction[3ex]{\langle \bar q_1 q_2 }{\bar q_3 }{q_4 | \bar q_5 b
\bar q_6 q_7 | }{\bar q_8 }
\langle \bar q_1 q_2 \bar q_3 q_4 | \bar q_5 b \bar q_6 q_7 | \bar q_8
b \rangle }$ &
$D= { \contraction[1ex]{\langle }{\bar q_1 }{}{q_2 }
\contraction[1ex]{\langle \bar q_1 q_2 \bar q_3 }{q_4 }{| \bar q_5 b
}{\bar q_6 }
\contraction[2ex]{\langle \bar q_1 q_2 \bar q_3 q_4 | }{\bar q_5 }{b
\bar q_6 q_7 | }{\bar q_8 }
\contraction[3ex]{\langle \bar q_1 q_2 }{\bar q_3 }{q_4 | \bar q_5 b
\bar q_6 }{q_7 }
\langle \bar q_1 q_2 \bar q_3 q_4 | \bar q_5 b \bar q_6 q_7 | \bar q_8
b \rangle }$ \\
$E= { \contraction[1ex]{\langle \bar q_1 q_2 \bar q_3 }{q_4 }{| \bar
q_5 b }{\bar q_6 }
\contraction[2ex]{\langle \bar q_1 q_2 }{\bar q_3 }{q_4 | \bar q_5 b
\bar q_6 }{q_7 }
\contraction[3ex]{\langle \bar q_1 }{q_2 }{\bar q_3 q_4 | }{\bar q_5 }
\contraction[4ex]{\langle }{\bar q_1 }{q_2 \bar q_3 q_4 | \bar q_5 b
\bar q_6 q_7 | }{\bar q_8}
\langle \bar q_1 q_2 \bar q_3 q_4 | \bar q_5 b \bar q_6 q_7 | \bar q_8
b \rangle }$ &
$F= { \contraction[1ex]{\langle \bar q_1 }{q_2 }{\bar q_3 q_4 | }{\bar
q_5 }
\contraction[2ex]{\langle \bar q_1 q_2 \bar q_3 }{q_4 }{| \bar q_5 b
}{\bar q_6 }
\contraction[3ex]{\langle \bar q_1 q_2 }{\bar q_3 }{q_4 | \bar q_5 b
\bar q_6 q_7 | }{\bar q_8 }
\contraction[4ex]{\langle }{\bar q_1 }{ q_2 \bar q_3 q_4 | \bar q_5 b
\bar q_6 }{q_7 }
\langle \bar q_1 q_2 \bar q_3 q_4 | \bar q_5 b \bar q_6 q_7 | \bar q_8
b \rangle }$ \\
$G= { \contraction[1ex]{\langle \bar q_1 }{q_2 }{}{\bar q_3 }
\contraction[1ex]{\langle \bar q_1 q_2 \bar q_3 }{q_4 }{| }{\bar q_5 }
\contraction[1ex]{\langle \bar q_1 q_2 \bar q_3 q_4 | \bar q_5 b
}{\bar q_6 }{}{q_7}
\contraction[2ex]{\langle }{\bar q_1 }{q_2 \bar q_3 q_4 | \bar q_5 b
\bar q_6 q_7 | }{\bar q_8}
\langle \bar q_1 q_2 \bar q_3 q_4 | \bar q_5 b \bar q_6 q_7 | \bar q_8
b \rangle }$ &
$H= { \contraction[1ex]{\langle \bar q_1 }{q_2 }{}{\bar q_3 }
\contraction[1ex]{\langle \bar q_1 q_2 \bar q_3 }{q_4 }{| }{\bar q_5 }
\contraction[1ex]{\langle \bar q_1 q_2 \bar q_3 q_4 | \bar q_5 b
}{\bar q_6 }{q_7 | }{\bar q_8 }
\contraction[2ex]{\langle }{\bar q_1 }{ q_2 \bar q_3 q_4 | \bar q_5 b
\bar q_6 }{q_7 }
\langle \bar q_1 q_2 \bar q_3 q_4 | \bar q_5 b \bar q_6 q_7 | \bar q_8
b \rangle }$ \\
$J= { \contraction[1ex]{\langle }{\bar q_1 }{}{q_2 }
\contraction[1ex]{\langle \bar q_1 q_2 }{\bar q_3 }{}{q_4 }
\contraction[2ex]{\langle \bar q_1 q_2 \bar q_3 q_4 | }{\bar q_5 }{b
\bar q_6 q_7 | }{\bar q_8 }
\contraction[1ex]{\langle \bar q_1 q_2 \bar q_3 q_4 | \bar q_5 b
}{\bar q_6 }{}{q_7}
\langle \bar q_1 q_2 \bar q_3 q_4 | \bar q_5 b \bar q_6 q_7 | \bar q_8
b \rangle }$ &
$K= { \contraction[1ex]{\langle }{\bar q_1 }{}{q_2 }
\contraction[1ex]{\langle \bar q_1 q_2 \bar q_3 }{q_4 }{| }{\bar q_5 }
\contraction[1ex]{\langle \bar q_1 q_2 \bar q_3 q_4 | \bar q_5 b
}{\bar q_6 }{}{q_7}
\contraction[2ex]{\langle \bar q_1 q_2 }{\bar q_3 }{q_4 | \bar q_5 b
\bar q_6 q_7 | }{\bar q_8 }
\langle \bar q_1 q_2 \bar q_3 q_4 | \bar q_5 b \bar q_6 q_7 | \bar q_8
b \rangle }$ \\
$ L= { \contraction[1ex]{\langle }{\bar q_1 }{}{q_2 }
\contraction[1ex]{\langle \bar q_1 q_2 \bar q_3 }{q_4 }{| }{\bar q_5 }
\contraction[1ex]{\langle \bar q_1 q_2 \bar q_3 q_4 | \bar q_5 b
}{\bar q_6 }{q_7 | }{\bar q_8 }
\contraction[2ex]{\langle \bar q_1 q_2 }{\bar q_3 }{q_4 | \bar q_5 b
\bar q_6 }{q_7 }
\langle \bar q_1 q_2 \bar q_3 q_4 | \bar q_5 b \bar q_6 q_7 | \bar q_8
b \rangle }$ & 
$N= { \contraction[1ex]{\langle \bar q_1 }{q_2 }{}{\bar q_3 }
\contraction[2ex]{\langle }{\bar q_1 }{ q_2 \bar q_3 }{q_4 }
\contraction[1ex]{\langle \bar q_1 q_2 \bar q_3 q_4 | \bar q_5 b
}{\bar q_6 }{}{q_7}
\contraction[2ex]{\langle \bar q_1 q_2 \bar q_3 q_4 | }{\bar q_5 }{b
\bar q_6 q_7 | }{\bar q_8 }
\langle \bar q_1 q_2 \bar q_3 q_4 | \bar q_5 b \bar q_6 q_7 | \bar q_8
b \rangle }$ \\
$S= { \contraction[1ex]{\langle \bar q_1 }{q_2 }{}{\bar q_3 }
\contraction[2ex]{\langle }{\bar q_1 }{ q_2 \bar q_3 }{q_4 }
\contraction[2ex]{\langle \bar q_1 q_2 \bar q_3 q_4 | }{\bar q_5 }{b
\bar q_6 }{q_7 }
\contraction[1ex]{\langle \bar q_1 q_2 \bar q_3 q_4 | \bar q_5 b
}{\bar q_6 }{q_7 | }{\bar q_8 }
\langle \bar q_1 q_2 \bar q_3 q_4 | \bar q_5 b \bar q_6 q_7 | \bar q_8
b \rangle }$ & 
$U= { \contraction[1ex]{\langle }{\bar q_1 }{}{q_2 }
\contraction[1ex]{\langle \bar q_1 q_2 }{\bar q_3 }{}{q_4 }
\contraction[2ex]{\langle \bar q_1 q_2 \bar q_3 q_4 | }{\bar q_5 }{b
\bar q_6 }{q_7 }
\contraction[1ex]{\langle \bar q_1 q_2 \bar q_3 q_4 | \bar q_5 b
}{\bar q_6 }{q_7 | }{\bar q_8 }
\langle \bar q_1 q_2 \bar q_3 q_4 | \bar q_5 b \bar q_6 q_7 | \bar q_8
b \rangle }$ \\
\end{tabular}
\end{center}
\label{contractions}
\end{table}

{}From here on, our goal is to minimize the number of independent Wick
contraction structures. We can do this simply by comparing them two
by two and using the following three rules to relate them:

(1) \textsl{Flavor symmetries:} Under isospin the contraction of two
    $u$ quarks is equivalent to that of two $d$ quarks; under flavor
    SU(3) symmetry this is true also for $s$ quarks. In the following,
    we always assume isospin symmetry. Note that since the effective
    hamiltonian is at leading order in the electroweak interaction,
    the addition of gluons does not violate isospin.

(2) \textsl{Fierz identities:} Since all operators have a $(V-A)\times
    (V-A)$ structure, the effect of a Fierz transformation is to
    simply exchange the first and the third quarks of the operators.
    As an example, we have
\beq 
{ \contraction[1ex]{\langle \bar u }{d }{}{\bar d }
\contraction[1.6ex]{\langle \bar u d \bar d }{s }{| \bar u b }{\bar s
} \contraction[2ex]{\langle \bar u d \bar d s | }{\bar u }{b \bar s
}{u } \contraction[3ex]{\langle }{\bar u }{d \bar d s | \bar u b \bar
s u | }{\bar u} \langle \bar u d \bar d s | \bar u b \bar s u | \bar u
b \rangle } = { \contraction[1ex]{\langle \bar u }{d }{}{\bar d }
\contraction[1.6ex]{\langle \bar u d \bar d }{s }{| }{\bar s }
\contraction[1.3ex]{\langle \bar u d \bar d s | \bar s b }{\bar u
}{}{u} \contraction[2ex]{\langle }{\bar u }{d \bar d s | \bar s b \bar
u u | }{\bar u} \langle \bar u d \bar d s | \bar s b \bar u u | \bar u
b \rangle }~,
\eeq
where the trivial contraction of $b$ fields is always understood. This
example shows that $A$-type and $G$-type contractions are related. In
general, all contractions are related in pairs.

(3) \textsl{Final-state symmetry:} In our notation, the order of
mesons in the final state is arbitrary. Thus, a change of this order ($\bar q_1 q_2 \leftrightarrow \bar q_3
q_4$) has no consequence. For example,
\beq 
{\contraction[0.6ex]{\langle \bar u }{d }{}{\bar d }
  \contraction[1ex]{\langle \bar u d \bar d }{d }{| \bar u b }{\bar d
  } \contraction[2ex]{\langle \bar u d \bar d d | }{\bar u }{b \bar d
  }{u } \contraction[3ex]{\langle }{\bar u }{d \bar d d | \bar u b
  \bar d u | }{\bar u} \langle \bar u d \bar d d | \bar u b \bar d u |
  \bar u b \rangle } = { \contraction[1ex]{\langle }{\bar d }{ d \bar
  u }{d } \contraction[1ex]{\langle \bar d d \bar u d | }{\bar u }{b
  \bar d }{u } \contraction[2.3ex]{\langle \bar d d }{\bar u }{d |
  \bar u b \bar d u | }{\bar u } \contraction[3ex]{\langle \bar d }{d
  }{\bar u d | \bar u b }{\bar d } \langle \bar d d \bar u d | \bar u
  b \bar d u | \bar u b \rangle }~,
\eeq 
showing that $A$-type and $Q$-type contractions are related. Again,
all contractions can be related in pairs.

Note that, although the above analysis is at the level of quarks
instead of mesons, one can prove that our results hold at the level of
mesons \cite{avenir}. The thrust of the proof is that we only compare
contractions two by two at the level of quarks, so that the mesons
affect both side of any equality identically. Thus, the equality
remains true at the level of mesons.

We begin by considering the tree contributions to $B\to \pi \pi$
decays. For $B^- \to \pi^- \pi^0$ we have (recall that isospin
symmetry is always assumed)
\beq 
T_{\sss \pi^- \pi^0} = \lambda_p^{(d)} c_i \langle \frac{1}{\sqrt{2}}
(\pi^- \pi^0 + \pi^0 \pi^- )|O^p_i | B^- \rangle~,
\eeq
where $p=u,c$ and a sum over $i=1,2$ is understood. Then,
\bea
T_{\sss \pi^- \pi^0} &=& \frac{\lambda_p^{(d)}}{2} c_i  (
\langle \bar u d \bar d d | (\bar p b \bar d p)_i | \bar u b \rangle 
-\langle \bar u d \bar u u | (\bar p b \bar d p)_i | \bar u b \rangle 
\nn\\
& &
+\langle \bar d d \bar u d | (\bar p b \bar d p)_i | \bar u b \rangle 
-\langle \bar u u \bar u d | (\bar p b \bar d p)_i | \bar u b \rangle 
)~.
\eea
The color indices are not written explicitly, but they are understood
with $i$ subscripts. When we sum over all Wick contractions and
simplify we then get
\beq
T_{\sss \pi^- \pi^0} =-\frac{\lambda_u^{(d)}}{2} c_i [I^u_i + M^u_i +
E^u_i + F^u_i]~,
\label{eqpipi1}\eeq
where the $u$ exponents stand for $p=u$. Using the final-state
symmetry [rule (3)] we have $c_i E^u_i = c_i M^u_i$ and $c_i F^u_i =
c_i I^u_i$, so that
\beq
T_{\sss \pi^- \pi^0} = -\lambda_u^{(d)}  c_i [E^u_i + F^u_i]~.
\eeq

A similar procedure can be carried out for $\bar B^0\to \pi^- \pi^+$
and $\bar B^0\to \pi^0 \pi^0$. We find
\bea
T_{\pi^- \pi^+} &=&-\lambda_u^{(d)} \sqrt{2} c_i [A^u_i + E^u_i +
H^u_i + S^u_i]\nn\\
& &-\lambda_c^{(d)} \sqrt{2} c_i [A^c_i + S^c_i]~,\nn\\
T_{\pi^0 \pi^0}&=&\lambda_u^{(d)} c_i  [A^u_i +H^u_i + S^u_i - F^u_i]\nn\\
& &+\lambda_c^{(d)} c_i [A^c_i + S^c_i]~.
\label{eqpipi2}
\eea

Comparing this parametrization in terms of contractions with that of
the language of diagrams \cite{GHLR} we can write all tree diagrams in
terms of Wick contractions:
\bea
T &=& \sqrt{2} c_i E_i^u~,\nn\\
C &=& \sqrt{2} c_i F_i^u~,\nn\\
E &=& \sqrt{2} c_i H_i^u~,\nn\\
P_{u,c} &=& \sqrt{2} c_i A_i^{u,c}~,\nn\\
PA_{u,c} &=& \sqrt{2} c_i S_i^{u,c}~,
\label{eqpipi4}\eea
where $T$ and $C$ are respectively the color-allowed and
color-suppressed tree diagrams, $E$ is the exchange diagram, and
$P_{u,c}$ and $PA_{u,c}$ are the tree parts which renormalize
respectively the gluonic penguin and penguin-annihilation amplitudes.
A priori there is an ambiguity in finding this one-to-one
correspondance, but this is completely removed by using the fact that
the exchange and the penguin-annihilation contributions cannot be
described by a contraction which contains a spectator quark ($E$-type
and $F$-type) and $T$, $C$ and $P_{u,c}$ must involve the spectator
quark. Note also that in this notation diagrams do not contain
Cabibbo-Kobayashi-Maskawa (CKM) factors.

For EWP's, the principle is exactly the same, but the operators are
slightly different. The result is
\bea
P^{\sss EW}_{\pi^- \pi^0}&=& \frac{3}{2} \lambda_t^{(d)} c_i [E_i +
F_i]~,\nn\\
P^{\sss EW}_{\pi^- \pi^+} &=& \frac{3}{2} \lambda_t^{(d)} \sqrt{2} c_i
[\frac{2}{3} F_i + \frac{1}{3} (B_i + G_i + N_i)\nn\\
& &- \frac{1}{3}(A_i + H_i + S_i)]~,\nn\\
P^{\sss EW}_{\pi^0 \pi^0} &=& \frac{3}{2} \lambda_t^{(d)} c_i [E_i +
\frac{1}{3} F_i - \frac{1}{3} (B_i + G_i + N_i)\nn\\
& &+ \frac{1}{3}(A_i + H_i + S_i)]~,
\label{eqpipi3}\eea
where $x^u_i = x^d_i \equiv x_i$ (with $x=A,B,C,...$) by isospin.
Contractions are of the form $\langle \bar q q \bar q q | \bar d b
\bar y y | \bar q b \rangle$ where the $q$'s are $u$ or $d$
independently, and $y=u,d$ (pieces with $y=s,c,b$ are absorbed into
gluonic penguin operators by isospin). A sum over $i=9,10$ is
understood. Again comparing with diagrams \cite{GHLR} we must have
\bea
P_{\sss EW} &=& - \frac{3}{2} \sqrt{2} c_i E_i~,\nn\\
P_{\sss EW}^c &=& -\frac{3}{2} \sqrt{2} c_i F_i~,
\label{diagpipi2}\eea
where we have omitted the other EWP's because they are not interesting
for our purpose. Comparing Eqs.~(\ref{eqpipi4}) and (\ref{diagpipi2}),
it is clear that $T$, $C$ and the EWP's are expressed in terms of the
same types of contractions. The order of quarks in their operators is
slightly different, but these are related by isospin [rule (1)].

We now have to explicitly compute the effect of the color indices.
For example,
\bea
\sum_{i=1,2} c_i E_i
&=&
c_1
{
\contraction[1ex]{\langle \bar q_{1x} q_{2x} \bar q_{3y} }{q_{4y} }{|
\bar q_{5i} b_i }{\bar q_{6j} }
\contraction[2ex]{\langle \bar q_{1x} q_{2x} }{\bar q_{3y} }{q_{4y} |
\bar q_{5i} b_i \bar q_{6j} }{q_{7j} }
\contraction[3ex]{\langle \bar q_{1x} }{q_{2x} }{\bar q_{3y} q_{4y} |
}{\bar q_{5i} }
\contraction[4ex]{\langle }{\bar q_{1x} }{q_{2x} \bar q_{3y} q_{4y} |
\bar q_{5i} b_i \bar q_{6j} q_{7j} | }{\bar q_{8z}} \langle \bar
q_{1x} q_{2x} \bar q_{3y} q_{4y} | \bar q_{5i} b_i \bar q_{6j} q_{7j}
| \bar q_{8z} b_z \rangle
}
\nn\\
& &+
c_2
{
\contraction[1ex]{\langle \bar q_{1x} q_{2x} \bar q_{3y} }{q_{4y} }{|
\bar q_{5i} b_j }{\bar q_{6j} }
\contraction[2ex]{\langle \bar q_{1x} q_{2x} }{\bar q_{3y} }{q_{4y} |
\bar q_{5i} b_j \bar q_{6j} }{q_{7j} }
\contraction[3ex]{\langle \bar q_{1x} }{q_{2x} }{\bar q_{3y} q_{4y} |
}{\bar q_{5i} }
\contraction[4ex]{\langle }{\bar q_{1x} }{q_{2x} \bar q_{3y} q_{4y} |
\bar q_{5i} b_j \bar q_{6j} q_{7i} | }{\bar q_{8z}} \langle \bar
q_{1x} q_{2x} \bar q_{3y} q_{4y} | \bar q_{5i} b_j \bar q_{6j} q_{7i}
| \bar q_{8z} b_z \rangle }~.~~
\eea
In the above, the first contraction is color-allowed, while the second
one is color-suppressed. Thus,
\bea
\sum_{i=1,2} c_i E_i
&=&
c_1 \delta_{xz} \delta_{xi} \delta_{yj} \delta_{zi}
{
\contraction[1ex]{\langle \bar q_1 q_2 \bar q_3 }{q_4 }{| \bar q_5 b
}{\bar q_6 }
\contraction[2ex]{\langle \bar q_1 q_2 }{\bar q_3 }{q_4 | \bar q_5 b
\bar q_6 }{q_7 }
\contraction[3ex]{\langle \bar q_1 }{q_2 }{\bar q_3 q_4 | }{\bar q_5 }
\contraction[4ex]{\langle }{\bar q_1 }{q_2 \bar q_3 q_4 | \bar q_5 b
\bar q_6 q_7 | }{\bar q_8} \langle \bar q_1 q_2 \bar q_3 q_4 | \bar
q_5 b \bar q_6 q_7 | \bar q_8 b \rangle }
\nn\\
& &+
c_2 \delta_{xz} \delta_{xi} \delta_{yi} \delta_{yj} \delta_{zj}
{
\contraction[1ex]{\langle \bar q_1 q_2 \bar q_3 }{q_4 }{| \bar q_5 b
}{\bar q_6 }
\contraction[2ex]{\langle \bar q_1 q_2 }{\bar q_3 }{q_4 | \bar q_5 b
\bar q_6 }{q_7 }
\contraction[3ex]{\langle \bar q_1 }{q_2 }{\bar q_3 q_4 | }{\bar q_5 }
\contraction[4ex]{\langle }{\bar q_1 }{q_2 \bar q_3 q_4 | \bar q_5 b
\bar q_6 q_7 | }{\bar q_8} \langle \bar q_1 q_2 \bar q_3 q_4 | \bar
q_5 b \bar q_6 q_7 | \bar q_8 b \rangle }\nn\\
&=& c_1 N_c^2 \bar E + c_2 N_c \bar E ~,
\eea
where the bar on $\bar E$ is added to stress on the fact that color
effects are extracted. 

Doing this for $T$, $C$, $E$, $P_{\sss EW}$ and $P_{\sss EW}^c$ from
Eqs.~(\ref{eqpipi4}) and (\ref{diagpipi2}) it is easy to find
\bea
T&=& \sqrt{2} (c_1 N_c^2 + c_2 N_c) \bar E~,\nn\\
C&=& \sqrt{2} (c_1 N_c + c_2 N_c^2) \bar F~,\nn\\
E&=& \sqrt{2} (c_1 N_c + c_2 N_c^2) \bar H~,\nn\\
P_{\sss EW} &=&  -\frac{3}{2} \sqrt{2} (c_9 N_c^2 +c_{10}N_c)\bar E~,\nn\\
P_{\sss EW}^c &=& - \frac{3}{2} \sqrt{2} (c_9 N_c +c_{10}N_c^2)\bar F~,
\label{diagpipi}\eea
which imply that for some specific ratios, the long-distance parts
(matrix elements) cancel:
\bea
\frac{P_{\sss EW}}{T} &=& - \frac{3}{2}
\frac{c_9 +\frac{c_{10}}{N_c}}{c_1 + \frac{c_2}{N_c}} \approx 0.013
~,\nn\\
\frac{P_{\sss EW}^c}{C} &=& -\frac{3}{2}
\frac{\frac{c_9}{N_c} + c_{10}}{\frac{c_1}{N_c}+c_2}\approx 0.013 ~.
\label{relpipi1}
\eea
These relations are new. For the Wilson coefficients, we have used
values given in Ref.~\cite{BBNS} evaluated at NLO with $\mu = m_b$.

But there is more. Using the final-state symmetry ($\bar E = \bar M$
and $\bar F = \bar I$), and adding Fierz transformations ($\bar E=\bar
I$ and $\bar F = \bar M$), we have $\bar E = \bar F$. This implies
\bea
\frac{C}{T} &=& \frac{\frac{c_1}{N_c}+c_2}{c_1
+ \frac{c_2}{N_c}}\approx 0.17~,\nn\\
\frac{P_{\sss EW}^c}{P_{\sss EW}} &=&
\frac{\frac{c_9}{N_c}+c_{10}}{c_9 + \frac{c_{10}}{N_c}}\approx 0.16~.
\label{relpipi2}
\eea
Again, these relations are new and confirm naive estimations of
color suppression \cite{GHLR}. Finally, Eqs.~(\ref{diagpipi}) also
imply
\bea
\frac{P_{\sss EW}^c + P_{\sss EW}}{C+T} &=&
-\frac{3}{2} \frac{\frac{c_9}{N_c} + c_{10} + c_9
+\frac{c_{10}}{N_c}}{\frac{c_1}{N_c}+c_2 + c_1 + \frac{c_2}{N_c}}
\nn\\
& = & -\frac{3}{2} \frac{c_9 + c_{10}}{c_1+c_2} \approx 0.013~,
\eea
which is exactly Eq.~(23) of Ref.~\cite{GPY} by Gronau, Pirjol and Yan
(GPY). However, note that with our approach the heavy formalism of
Clebsch-Gordan coefficients is not required. This result also
represents a cross-check of our calculation.

A similar exercise can be carried out for $B\to \pi K$ decays. For
simplicity, we use the same notation as for $B\to \pi \pi$, but it is
clear that, for example, an A-type contraction in $B\to \pi \pi$ is
not the same as that in $B\to \pi K$ because of the $s$-quark fields,
unless we assume flavor SU(3) symmetry. To keep track of different
flavors, primed contractions have the form $\langle \bar q q \bar q
s|\bar s b \bar q q|\bar q b\rangle$ and non-primed contractions have
the form $\langle \bar q q \bar q s|\bar q b \bar s q|\bar q b\rangle$
($q=u,d$ independently). Note also that, contrary to $B\to \pi \pi$,
we have no symmetry in the final state and this leads to some
differences. Again we can express the graphical amplitudes of
Refs.~\cite{GHLR} in terms of our contractions:
\bea
T'&=& \sqrt{2} (c_1 N_c^2 + c_2 N_c) \bar E~,\nn\\
C'&=& \sqrt{2} (c_1 N_c + c_2 N_c^2) \bar F~,\nn\\
A'&=& \sqrt{2} (c_1 N_c^2 + c_2 N_c) \bar B~,\nn\\
P'_{\sss EW} &=& - \frac{3}{2} \sqrt{2} (c_9 N_c^2 +c_{10}N_c)\bar
M'~,\nn\\
P_{\sss EW}^{'c} &=& - \frac{3}{2} \sqrt{2} (c_9 N_c +c_{10}N_c^2)\bar
I'~,
\label{diagpik}
\eea
where $A'$ is the annihilation amplitude. We have omitted the other
diagrams since they are not interesting here. To go further, we use
Fierz transformations, which give $\bar E = \bar I'$ and $\bar F =
\bar M'$. We then obtain
\bea
\frac{P_{\sss EW}^{'c}}{T'} &=& - \frac{3}{2}
\frac{\frac{c_9}{N_c}+c_{10}}{c_1 + \frac{c_2}{N_c}}\approx 0.0022
~,\nn\\
\frac{P'_{\sss EW}}{C'} &=& - \frac{3}{2}
\frac{c_9+\frac{c_{10}}{N_c}}{\frac{c_1}{N_c}+c_2}\approx 0.08 ~.
\label{relpik}
\eea
Again, these two relations are new. The difference between these
equations and those of Eqs.~(\ref{relpipi1}) is due to the absence of
symmetry in the final state. These two equations are important since
it is generally believed that it is impossible to relate tree diagrams
and EWP's without SU(3) symmetry (for example, see Ref.~\cite{nosu3}).

We have stressed on several occasions that all previous relations do
not require flavor SU(3). However, if we add this symmetry, the
position of the $s$ quark is no longer important, so that there is a
final-state symmetry in $B \to \pi K$. Indeed, this decay can be
related to $B\to \pi \pi$. Under SU(3) we have $\bar E = \bar F = \bar
M' = \bar I'$. It is then easy to show that Eqs.~(\ref{relpipi2}) are
valid also for $B\to \pi K$. In addition, using the final-state
symmetry in Eqs.~(\ref{diagpik}), one can derive
\beq
\frac{P_{\sss EW}^{'c} + P'_{\sss EW}}{C'+T'} =
- \frac{3}{2} \frac{c_9 + c_{10}}{c_1+c_2} \approx 0.013 ~,
\eeq
which is the well known Neubert-Rosner relation \cite{NR, GPY}. Also,
one can easily derive (it is long but straightforward) the relation
\bea
P^{\sss EW}_{\pi^+ K^-} + P^{\sss EW}_{\pi^0 \bar K^0} &=& \frac{3}{4}
\frac{c_9-c_{10}}{c_1-c_2}(A+C-T-E)\nn\\
&-& \frac{3}{4} \frac{c_9+c_{10}}{c_1+c_2}(A-C-T+E)~,~~
\label{longcheck}
\eea
which is exactly the CP-conjugate of Eq.~(18) of GPY \cite{GPY}. The
fact that we reproduce known results \cite{NR,GPY} is an important
cross-check to our calculation. Indeed, the reproduction of
Eq.~(\ref{longcheck}) above is particularly important since this
equation is quite complicated. At every step our calculation is
consistent with all known SU(3) relations. Note that it supports the
fact that we are working at the level of mesons since GPY are clearly
working at this level.

Finally, there is an interesting relation involving $A$ and $E$
diagrams. From Eqs.~(\ref{diagpipi}) and (\ref{diagpik}), we have
\beq
\frac{E}{A} = \frac{\frac{c_1}{N_c}+c_2}{c_1
  +\frac{c_2}{N_c}}\approx 0.17~.
\eeq
Again, this relation is new. 

Before concluding, there is an important issue we must address. We
have derived several new relations among diagrams. Why could these new
relations not be obtained from standard systematic isospin and SU(3)
analysis?  The answer is simple. Consider tree diagrams for
example. There are six tree topologies: $T$, $C$, $A$, $E$, $P_u$ and
$PA_u$. However, under the SU(3) formalism, only five linear
combinations of these six topologies can appear in any amplitude:
$P_u+T$, $P_u+A$, $C-P_u$, $P_u+PA_u$ and $C-E$.  Consequently, it is
possible to find relations among these five linear combinations of
diagrams. However, it is impossible to find a relation between $T$ and
$C$ alone, for example. This is because $T$ and $C$ alone do not exist
in this formalism. On the other hand in our approach, isolated
diagrams are well-defined contractions. Relations among contractions
automatically imply relations among diagrams.

To summarize, the neglect of the $(V-A)\times (V+A)$ pieces of the
electroweak penguin (EWP) amplitudes in the effective hamiltonian
allows us to describe both tree and EWP operators in terms of
$(V-A)\times (V-A)$ interactions. By computing the Wick contractions
of these operators sandwiched between initial and final states, we are
able to make the connection between diagrams in $B$ decays and these
contractions. The ratios of the sizes of various diagrams can then be
expressed as a ratio of contractions. Note that these contractions
include (uncalculable) matrix elements. However, the key point is
that, in certain ratios, the matrix elements cancel due to symmetry
arguments, so that the ratio of sizes of diagrams is expressible
purely in terms of (calculable) Wilson coefficients. For the case
where this symmetry is purely isospin, we have presented a variety of
new results in $B\to \pi \pi$ and $B \to \pi K$ decays. These are
rigorous, and are consistent with naive estimates \cite{GHLR}. If the
symmetry is extended to flavor SU(3), we get additional results, all
in agreement with published papers (especially Refs.~\cite{GPY, NR}).

The potential for applications of this method is great in $B$ and $D$
decays. For example, methods for extracting CKM weak phases from $B\to
\pi \pi$ and $B\to\pi K$ decays can be greatly improved. The standard
model can be tested in $B\to \pi \pi$ measurements and $\gamma$ can be
extracted from $B\to \pi K$ decays without using SU(3) approximations.
Also, estimates of SU(3) breaking should be facilitated since we can
avoid the heavy formalism of Clebsch-Gordan coefficients in our
approach. Applications, as well as explicit calculations, are
discussed in more detail in Ref.~\cite{avenir}.

\vskip3truemm 
I wish to thank D. London, Th. Mannel and Th. Feldmann for useful discussions.  I especially thank D. London for his help with the manuscript.  Finally, i thank everyone from LAPTH
in France and Universit\"at Siegen in Germany for their great
hospitality. This work was financially supported by NSERC of Canada.

\end{document}